\documentclass[oneside,english,twocolumn,tighten,letter,latin9]{aa}
\usepackage[T1]{fontenc}
\usepackage{textcomp}
\usepackage{babel}
\usepackage{graphicx}
\usepackage[]
 {hyperref}
\begin{document}
\title{What halts the growth of galaxies?}
\abstract{The gas reservoir of galaxies can be altered by outflows driven by
star-formation and luminous active galactic nuclei. Jets heating the
surroundings of host galaxies can also prevent gas cooling and inflows.
Spectacular examples for these three mass displacement channels have
been observed, but their importance in transforming the galaxy population
depends on the occurrence rates of outflow triggers. We aim to investigate
the absolute and relative importance of these three channels. In an
observation-driven approach, we combine distribution functions and
scaling relations to empirically compare average outflow rates across
the galaxy total stellar mass spectrum and across cosmic time. This
hinges on local outflow studies which should be extended to systematic,
large and diverse samples, and we do not yet consider a halo heating
effect by radiation-driven outflows. Our results show, independent
of simulations, the dominance of star formation-driven outflows in
low-mass galaxies. Massive galaxies today are predominately prevented
from growing further by jet heating, while at $z=1-3$ all three processes
are approximately similarly important. Over the full mass spectrum
and cosmic history, outflows driven by the radiation from active galactic
nuclei is never the dominant process.}
\author{Johannes Buchner\thanks{\protect\href{mailto:johannes.buchner.acad@gmx.com}{johannes.buchner.acad@gmx.com}}}
\institute{Max Planck Institute for Extraterrestrial Physics, Giessenbachstrasse,
85741 Garching, Germany}

\maketitle

\section{Introduction}

Star formation (SF) and the growth of super-massive black holes (SMBH)
are processes thought to shape the evolution of galaxies. Evidence
for this comes from cosmological simulations, where such feedback
mechanisms are effective in reproducing e.g., the luminosity function
and color distribution of galaxies \citep[e.g.][]{Croton2006}. However,
the observational evidence on the role and importance of the various
feedback mechanisms is more complex.

\begin{figure*}
\begin{centering}
\includegraphics[width=0.9\textwidth]{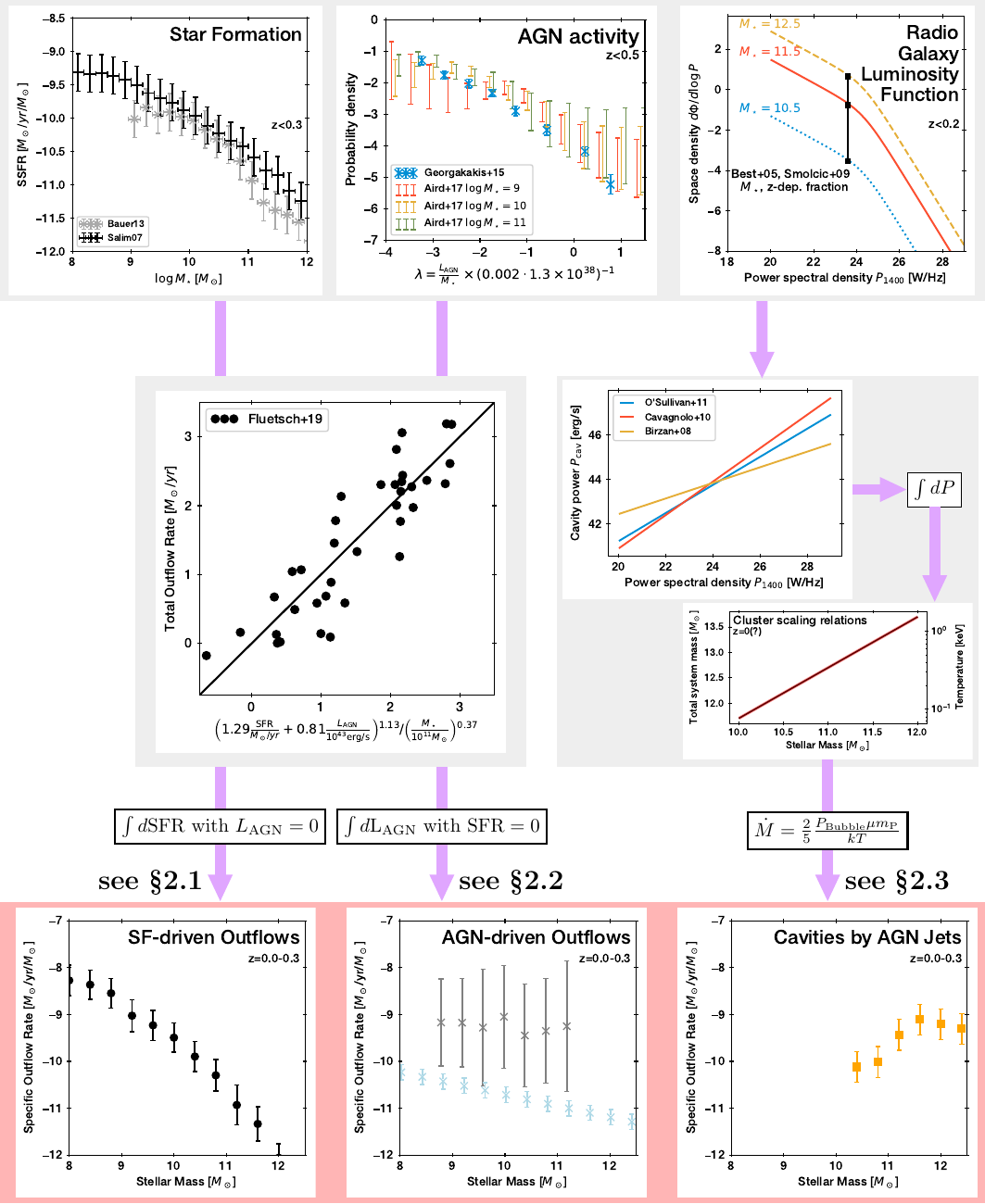}
\par\end{centering}
\caption{\protect\label{fig:outflowcalc}Our workflow starts from distribution
functions (top row), which are convolved with scaling relations (middle)
into rates of mass displacement (bottom row) at a given galaxy stellar
mass and redshift. Shown here is the lowest redshift bin.}
\end{figure*}

Observations are most clear in the rare extremes: very massive systems
where jets produced near the SMBH heat intergalactic matter \citep[e.g. see reviews of][]{Fabian2012,Heckman2014},
and bulk outflows of gas associated with rapid SMBH growth as an Active
Galactic Nucleus (AGN) \citep[e.g.,][]{Cicone2014}. In most systems,
disentangling the proposed feedback mechanisms and establishing causality
is observationally difficult. This is in part due to the different
time-scales AGN and SF act on in comparison to the impact of their
activity \citep[e.g.][]{Hickox2014}. Nevertheless, multi-variate
analyses of local galaxies suggest that the outflow rates are a function
of AGN luminosity, SFR and $M_{\star}$ \citep[e.g.][]{Fluetsch2018}.
To circumvent peculiarities of individual systems, large surveys of
galaxies have been used to establish statistical differences, for
example in AGN and non-AGN host galaxies. Understanding the selection
effects is crucial however, as are the limitations in characterizing
the host galaxy when it is contaminated with AGN light. Nevertheless,
the AGN accretion rate distributions as a function of basic host galaxy
properties such as stellar mass ($M_{\star}$) and star formation
rate (SFR) over cosmic time has been measured \citep[e.g.,][]{Aird2012,Bongiorno2012,Aird2018,Georgakakis2017a}.
Even more robustly established is the feedback in galaxy clusters
by linking AGN luminosity and cluster heating (see below).

In this work, we combine existing measurements to quantify which mechanisms
are effective at carrying gas out of a galaxy. We focus on outflows
associated with SF activity, AGN jets and AGN luminosity, and investigate
their importance as a function of galaxy mass and cosmic time.

\section{Methodology}

The goal of this work is to compute the average outflow rates of different
feedback mechanisms as a function of stellar mass and redshift.

For SF and AGN-related outflows, we rely on the relation of \citet[F19 hereafter]{Fluetsch2018}.
F19 found that the total (molecular, ionized and neutral hydrogen)
outflow mass rate can be accurately predicted given stellar mass,
SFR and AGN luminosity (see left middle panel of Figure~\ref{fig:outflowcalc}):

\begin{equation}
\dot{M}/M_{\odot}=(1.29\times\mathrm{SFR}+0.81\times L_{\mathrm{AGN},43})^{1.13}/M_{\star,11}^{0.37}\label{eq:fluetsch}
\end{equation}
where $L_{\mathrm{AGN,43}}$ is the bolometric AGN luminosity in units
of $10^{43}\mathrm{erg}/s$, $M_{\star,11}$ is the stellar mass in
units of $10^{11}M_{\odot}$, SFR is the star formation rate in units
of $M_{\odot}/\mathrm{yr}$. 

\subsection{Outflows from star formation}

\label{subsec:SFOutflows}We obtain (specific) SFR distributions \citep{Whitaker2014,Tomczak2016,Smit2014,Schreiber2015,Salmon2015,Salim2007,McLure2011,Labbe2013,Karim2011,Kajisawa2010,Bauer2013,Zwart2014}
from the literature compilation of \citet{Behroozi2018}. For each
of these SFR distributions measured at a certain stellar mass and
redshift (shown in top left panel of Figure~\ref{fig:outflowcalc}
for $z=0$), we use the F19 relation with $L_{\mathrm{AGN},43}=0$
to compute the outflow rate due to SFR alone. Integration over the
SFR probability distribution gives the average outflow rate at a given
stellar mass (bottom left panel of Figure~\ref{fig:outflowcalc}).
We use the variance across measurements as approximate uncertainties.

\begin{figure*}
\begin{centering}
\includegraphics[width=1\textwidth]{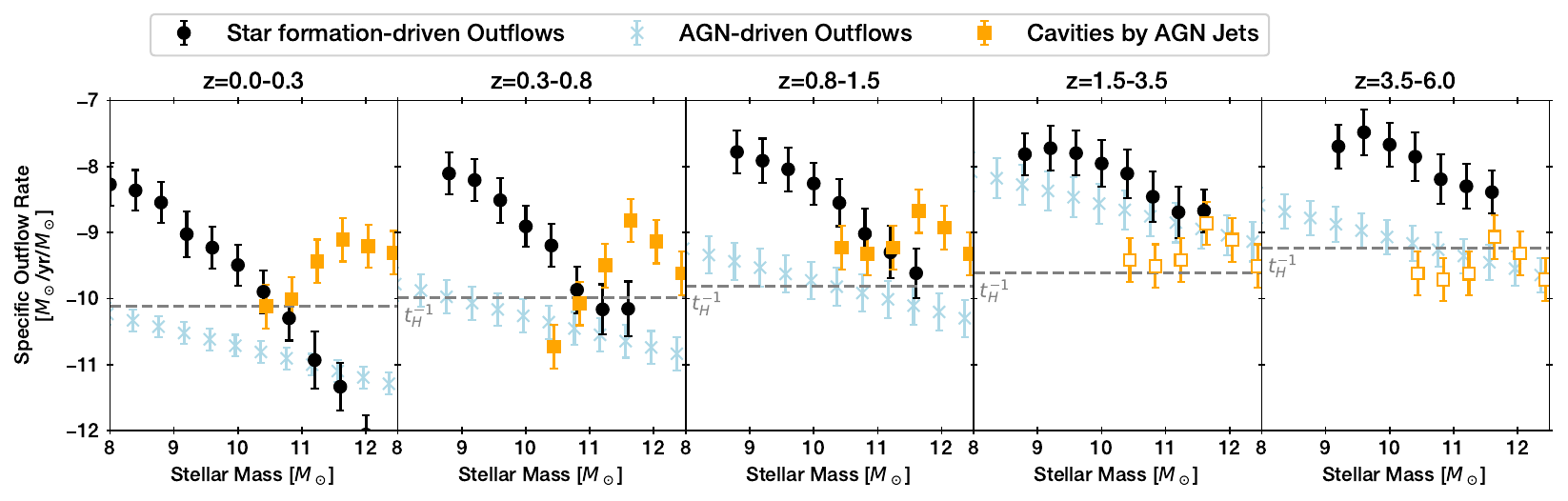}
\par\end{centering}
\caption{\protect\label{fig:Mass-dep-outflows}Mass dependence of outflow channels.
Panels indicate different redshift intervals. SF-driven outflows (black
error bars, §\ref{subsec:SFOutflows}), AGN-driven outflows (blue,
§\ref{subsec:AGNOutflows}) and AGN jet cavities preventing inflows
(orange, §\ref{subsec:AGNbubbles}) are compared. AGN outflows are
only weakly dependent of stellar mass. The horizontal dashed line
indicates the inverse Hubble time at that redshift. In the two right-most
redshift panels, the AGN jet power has been extrapolated (open orange
points).}
\end{figure*}

\subsection{Outflows radiatively pushed by active galactic nuclei}

\label{subsec:AGNOutflows}To compute the outflows associated with
luminous AGN, we require the AGN luminosity as a function of stellar
mass. \citet{Aird2018} found a power-law specific accretion rate
distribution (SARD), $p(\lambda|M_{\star})\propto L_{\mathrm{AGN}}/M_{\star}$
over a range of stellar masses and redshifts. At low redshifts, these
measurements are shown in the top middle panel of Figure~\ref{fig:outflowcalc}.
They start from a galaxy sample and compute for each galaxy the AGN
luminosity (from X-ray emission) and stellar mass (from broad-band
photometry). However, because of limited survey area, their sample
is not sensitive to constraining the bright end of the luminosity
function (near $\lambda\geq1$). There, the uncertainties are large
and follow a Gaussian extrapolation, because they assume a two-component
Gaussian mixture model to empirically fit the distribution. \citet{Georgakakis2017a}
uses also large-area surveys, however their methodology is different.
Instead of starting with a galaxy sample, they detect a AGN sample
in the X-rays and by SED fitting measure the SARD of host galaxies.
They also do not assume a SARD shape, but estimate the space density
in redshift and SARD bins. They find a universal SARD shape corresponding
to a powerlaw with index $-1$, with an exponential cut-off at approximately
twice the Eddington limit. Such a limit is also required to match
the SARD with the X-ray luminosity function \citep{Aird2013}. Here
we generally use the measurements of \citet{Aird2018}, and fully
propagate the space density uncertainties. When uncertainties become
large (>0.5dex) at the bright end, we impose a exponential cut-off
of \citet{Georgakakis2017a} to the upper error bars of \citet{Aird2018}.
We propagate the uncertainties in the relative normalisations. Our
combination, which is very similar to the recent results of \citet{Laloux2024},
is shown as the dashed error bars in the top middle panel of Figure~\ref{fig:outflowcalc}.

After converting the AGN luminosity to an outflow rate with eq.~\ref{eq:fluetsch}
(setting SFR=0), we integrate over the SARD to obtain total outflow
rate at a given stellar mass. The outflow rates are shown in the bottom
middle panel of Figure~\ref{fig:outflowcalc}.

\subsection{Bubbles blown by active galactic nuclei}

\label{subsec:AGNbubbles}In the centers of galaxy clusters and massive
galaxies, radio-emitting Active Galactic Nuclei are common \citep[e.g.,][]{Sabater2019}.
Their radio jets are thought to inject heat into the cluster gas that
would otherwise cool quickly, and the created hot bubbles rise slowly
by boyancy and slowly disperse their energy \citep[see e.g.,][]{Fabian2012}.
Outside gas that would otherwise fall into the cluster, condense onto
and thereby grow a galaxy, is stalled at large distances \citep[see e.g.,][]{Croton2006}.
To compute the mass accretion prevented by this mechanism, we combine
several correlations.

We start with the AGN radio luminosity function shape determined by
\citet{Sadler2002}. Comparison of radio luminosity works find that
the shape shows little evolution and is consistent between works and
radio wavelengths \citep[e.g.][]{Smolvcic2009,Heckman2014,Sabater2019a}.
The fraction of AGN radio-emitting more than $4\times10^{23}\mathrm{W}/\mathrm{Hz}$
is mass dependent \citep[their figure 11]{Smolvcic2009}. The combination
is a mass-dependent radio luminosity function, shown in the top right
panel of Figure~\ref{fig:outflowcalc}. Next, the radio luminosity
of AGN correlates well with the power stored in bubbles \citep[e.g.][]{Birzan2008,Cavagnolo2010,OSullivan2011}.
This is shown in the second plot in the right column. Integrating
over the power distribution, we estimate the total power injected
into bubbles by AGN jets. The bubble power is eventually radiatively
cooled away by cooling flows. The accretion rate is \citep{Fabian1994a}:
\begin{equation}
\dot{M}=\frac{2}{5}\frac{P_{\mathrm{Bubble}}\times\mu m_{\mathrm{P}}}{kT}\label{eq:coolingflow}
\end{equation}
Here the bubble power is used as the cluster luminosity to be emitted
at plasma temperature $T$. The remaining constants are the Boltzmann
constant $k$, the proton mass $m_{\mathrm{P}}$ and the mean molecular
mass, $\mu\approx0.6$ in ionized plasmas \citep{Pratt2019}. As cluster
temperature $T$ we use the relatively tight relation between stellar
mass and total mass of the system \citep{Vikhlinin2009}, combined
with the relation of total system mass to temperature \citep{Gonzalez2013,Kravtsov2018}.
These relations are shown in the left and right axes of the third
panel in the right column of Figure~\ref{fig:outflowcalc}.

The resulting total mass rate is shown in the bottom right panel of
Figure~\ref{fig:outflowcalc}. We propagate the systematic uncertainties
of the cavity-radio power relations, which vary most between works,
and the fit uncertainties in the mass-temperature and $M_{500}-M_{\star}$
relation. Unlike the outflows rates in the previous two sections,
eq~\ref{eq:coolingflow} computes the rate of an inflow that could
have happened, were it not prevented by the AGN energy injection.

\section{Results}

Our main result is presented in Figure~\ref{fig:Mass-dep-outflows}.
The importance of SF-driven outflows (black, §\ref{subsec:SFOutflows}),
AGN-driven outflows (blue, §\ref{subsec:AGNOutflows}) and AGN jet
bubbles preventing inflows (orange, §\ref{subsec:AGNbubbles}) are
compared for different redshift intervals. Points above the dashed
horizontal line indicate that the process is important, as the outflow
rate exceeds the current stellar mass divided by Hubble time. The
main findings are: (1) In massive systems, AGN jets dominate. (2)
In low-mass systems SF-driven outflows dominate. (3) At higher redshifts
($z\sim2$), all three channels appear to be important.

\begin{figure}
\begin{centering}
\includegraphics[width=1\columnwidth]{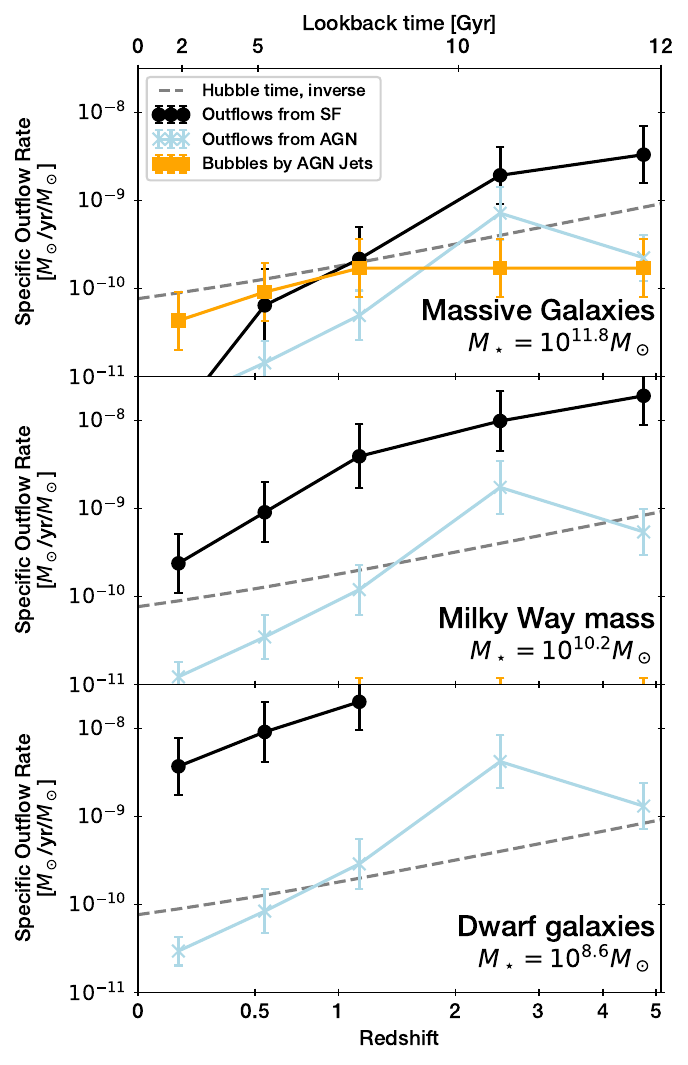}
\par\end{centering}
\caption{\protect\label{fig:Redshift-evolution}Outflow rate redshift evolution
in three mass bins. The lower mass systems (lower panels) are dominated
by SF outflows (black), while in the high-mass systems (top panel)
at low redshift, AGN jet bubbles (orange) dominate.}
\end{figure}

Figure~\ref{fig:Redshift-evolution} presents the same information
as a function of redshift. For massive galaxies (top panel), SF-driven
outflows was most important at $z>1$, but at present time AGN jets
are the dominant channel. For smaller systems (middle and bottom panel),
SF-driven outflows always dominate. This figure should be interpreted
with care, as galaxies grow in mass over cosmic time and thus do not
remain in a single panel.

\section{Discussion and conclusion}

We presented a framework to compare key channels influencing the growth
of galaxies. For this purpose, we combined state-of-the-art distribution
functions and scaling relations of outflows driven by SF or AGN and
cavities blown by AGN jets. Our approach is analogous to semi-analytic
modelers treatment of gas reservoirs that are filled by gravitational
collapse. The gas reservoir can be emptied by outflows and gas accretion
prevented by the heating of the galaxy intergalactic medium \citep[see e.g.,][]{Croton2006}.

In our work, we have however not relied upon simulations. Instead,
we obtained our results by combining observed relations collated from
a large body of research undertaken over the last decades. As a prediction
from Figure~\ref{fig:Redshift-evolution}, we should see AGN-driven
outflows dominate are widespread at $z\sim1-2$, and at high galaxy
stellar masses they dominate over SF-driven outflows. This is what
is indeed found in high-resolution integral-field observations \citep[e.g.][]{Genzel2014a}.

We acknowledge that several of the used relations have considerable
caveats. To name but a few examples: A more accurate determination
of the AGN activity in the luminous regime is required to accurately
determine the total luminosity density (see large gray error bars
in the middle bottom panel of Figure~\ref{fig:outflowcalc}). We
have extrapolated a local outflow relation out to high redshifts and
across a wide mass range. The Fluetsch relation needs to be extended
to larger, well-selected samples. As discussed in §4.2 of the recent
review by \citet{Harrison2024}, the scatter in the relation may be
very large due to a large fraction of luminous AGN lacking strong
outflows \citep[e.g.,][]{RamosAlmeida2022}. However, because the
population calculation in a given AGN luminosity and stellar mass
bin requires the (arithmetic) mean outflow rate, the change is potentially
small, and even if not would further reduce the low total population
outflow rate we find, not altering our conclusions. While the relation
was established at moderate to high-mass galaxies, low mass galaxies
are more frequently satellite galaxies that may undergo additional
environmental quenching \citep[see e.g.,][]{Peng2010,Bluck2020}.
Such additional quenching mechanisms are not considered here. More
work is also needed to understand the validity of applying cluster
scaling relations down to the galaxy group regime, and the importance
of projection effects in cavity energy relations \citep{Eckert2021}.

We focused on AGN-driven gas ejection because it is a common narrative.
Nevertheless, feedback by radiation-driven feedback of AGN may be
effective differently. For example, the energy injected may induce
a similar halo heating effect as radio jets, as suggested in simulations
\citep[e.g.,][]{Gabor2014,Bower2017,Nelson2019} and evidenced in
hot halo gas found by the Sunyaev--Zel\textquoteright dovich effect
in quasars \citep[e.g.,][]{Ruan2015,Crichton2016,Jones2023}. To address
this, the presented framework could be extended in the future with
a hybrid of the radio jet heating and AGN outflow calculations. Indeed,
a similar work to ours was recently conducted by \citet{Heckman2023},
which came to similar conclusions. While this work focused on mass
outflow rate, \citet{Heckman2023} focused on energy injection, which
can address the entropy injected by AGN outflows as an alternative
mechanism. Their approach to feedback from massive stars relied on
models, while our approach relies on observed outflow rate relations.
Both approaches have systematic uncertainties \citep[see e.g.,][]{Hardcastle2020}.
Their jet bubble calculations are similar to ours, but we calculate
from there the prevented rate of gas inflow. For AGN radiation-driven
outflows, we use the more recent tight relation by \citet{Fluetsch2018},
which allows disentangling causation by star formation and AGN. That
the two works with different approaches reach the same conclusions
is reassuring. The reverse cross-over case, where equatorial radio
jets drive molecular outflows \citep[see discussion in §4.2 of][]{Harrison2024}
could also be considered in the future, but would require further
observational studies that establish relations.

In summary, this study finds that transportation of gas out of galaxies
driven by luminous AGN is not an effective feedback mechanism. SF-driven
outflows are effective in removing gas in the low-mass regime, while
AGN jets are effective in the high-mass regime. Unless there are order-of-magnitude
under-estimates in our assumptions, outflows driven by AGN radiation
that extract gas from galaxies and is thus unavailable for star formation
is a sub-dominant galaxy evolution process. Nevertheless, in massive
galaxies and especially at $z=1-3$, the outflow rates of all three
channels considered in this work are comparable and of almost equal
importance. 
\begin{acknowledgements}
JB thanks Chris Harrison, Dominique Eckert, Andrea Merloni, Antonis
Georgakakis, Mara Salvato and Cristina Ramos Almeida for insightful
conversations and feedback on this manuscript. JB thanks the anonymous
referee for their helpful comments.
\end{acknowledgements}

\bibliographystyle{aa}
\bibliography{ref}

\end{document}